\documentclass{article}
\usepackage{bibentry}
\usepackage[authoryear]{natbib}
\usepackage{geometry}
\usepackage{enumitem}
\usepackage{hyperref}           
\usepackage{microtype}  	
\usepackage{graphicx}
\usepackage{amsfonts, amsmath, amsthm, amssymb}
\usepackage{float}

\begin{document}

\vspace{4em}
\noindent
\Large\textbf{Dissociating model architectures from inference computations} 
\\

\small{
\noindent
\textbf{Noor Sajid} \\
Kempner Institute for the Study of Natural and Artificial Intelligence, Harvard University, Cambridge, USA; \\
Max Planck Institute for Biological Cybernetics, Tübingen, Germany \\
noorsajidt@gmail.com

\vspace{0.5em}
\noindent
\textbf{Johan Medrano} \\
Lisa Yang Integrative Computational Neuroscience Center, MIT, Cambridge, MA, USA; \\ 
Department of Biological Engineering, MIT, Cambridge, MA, USA; \\
Functional Imaging Laboratory, Queen Square Institute of Neurology, UCL, London, UK \\
johmedr0@mit.edu}

\vspace{3em}
\noindent\rule{\textwidth}{0.6pt}
\normalsize



\paragraph{Abstract}

\cite{parr2025beyond} examines how auto-regressive and deep temporal models differ in their treatment of non-Markovian sequence modelling. Building on this, we highlight the need for dissociating model architectures---i.e., how the predictive distribution factorises---from the computations invoked at inference. We demonstrate that deep temporal computations are mimicked by autoregressive models by structuring context access during iterative inference. Using a transformer trained on next-token prediction, we show that inducing hierarchical temporal factorisation during iterative inference maintains predictive capacity while instantiating fewer computations. This emphasises that processes for constructing and refining predictions are not necessarily bound to their underlying model architectures. 


\vspace{2em}
\noindent\rule{\textwidth}{0.6pt}
\vspace{1em}
\subsection*{Introduction}
Transformers are remarkably good at integrating information across long sequences, despite lacking an explicit temporal hierarchy. \cite{parr2025beyond} show that attention mechanisms enable transformers to approximate non-Markovian inference by selectively attending to past inputs embedded in a continuous latent space. To understand why this matters, it is helpful to revisit what it means for a process to be non-Markovian. A system is non-Markovian if the current state fails to contain all the information necessary to predict the future, i.e., past context contributes to current predictions. This non-Markovianity often arises from partial observability of a higher-dimensional Markovian system~\citep{parr2025beyond}. Even local neural circuits can seem non-Markovian in isolation, even though they are embedded in distributed networks that integrate and maintain latent contexts~\citep{chaudhuri2015large,huth2016natural}.

To understand how systems manage such dependencies, it is useful to distinguish between: $i)$ the model architecture, which constrains the temporal dependencies a model can, in principle, represent, and $ii)$ the inference-time computations by which predictions are generated and refined. Deep temporal models address this by explicitly instantiating a hierarchy of latent variables and using Bayesian inference to propagate information up and down this hierarchy~\citep{friston2018deep,friston2020generative,parr2021generative}. In such hierarchies, stacking layers whose integration windows double at each stage gives an exponentially expanding receptive field, thereby multiplying effective memory capacity with each added level~\citep{kiebel2008hierarchy}. Such architectures have been used to effectively characterise functional hierarchies in the brain~\citep{friston2018deep,medrano2024linking}, including domains like language~\citep{friston2020generative}, vision~\citep{parr2021generative}, and motor control~\citep{yuan2023hierarchical}. 

Here, we claim that transformers can emulate these hierarchical computations without architectural modification. We show that by imposing a hierarchical pattern of context access during iterative inference—selectively querying past tokens at exponentially increasing intervals—an unmodified transformer can emulate the efficient, multi-timescale computations of deep temporal hierarchies.

\begin{figure}[!t]
    \centering
    \includegraphics[width=\linewidth]{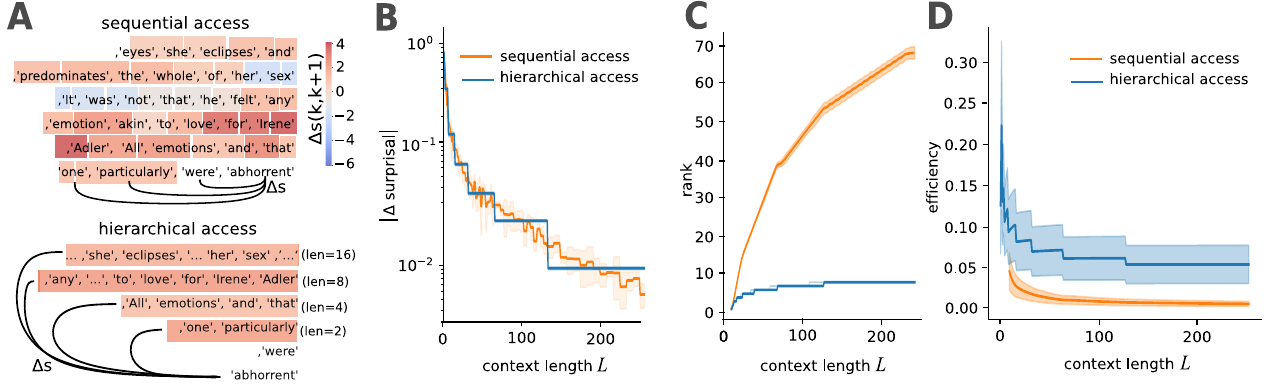}
    \caption{Comparison of sequential and hierarchical context access in language models predicting text from The Adventures of Sherlock Holmes~\citep{doyle1992adventures}. \textbf{A.} Schematic illustrating differences in context access scheme: the sequential path (top) accumulates tokens linearly, while the hierarchical path (bottom) samples context at exponentially increasing intervals (power law of order $2$). 
    \textbf{B.} show the absolute change of surprisal under each context access scheme, \textbf{C.} the rank (i.e., number of linearly independent rows/columns) of the context-surprisal transition matrix, and \textbf{D.} the predictive efficiency defined as the ratio of rank to the amplitude of change of surprisal, as a function of context length ($0–250$ units) across sequential (orange) and hierarchical (blue) processing. Solid lines denote means; shaded areas indicate standard error.}
    \label{fig:results}
\end{figure}

\subsection*{Methods}
To support our claim, we use a transformer model for language generation to numerically compare surprisal changes under two different token accumulation schemes: ($i$) a sequential scheme, which increases context length linearly and ($ii$) a hierarchical accumulation scheme, by accumulating token sets whose length equals the current context length (Fig.~\ref{fig:results}A). 
This hierarchical accumulation reproduces the inductive bias that ties hierarchical and temporal depth in deep temporal models~\citep{yu2015multi}, effectively doubling the context length or receptive field with each additional hierarchical level. 

\paragraph*{Experimental details} 
We use the (\href{https://huggingface.co/google/gemma-2b}{Gemma-2B}) transformer language model to compute token-level surprisal on The Adventures of Sherlock Holmes~\citep{doyle1992adventures} text. The surprisal of target token $w_t \in W$ under context length $k \in \{1,…,250\}$ is $S_k (w_t)= -\ln p(w_t|w_{t-1}, \cdots, w_{t-k})$. We define the change of surprisal as $\Delta S^{\rm \, seq.}_k(w_t)  = S_{k} (w_t) - S_{k-1} (w_t) = - \ln p(w_t|w_{t-k})$ in the sequential access case, and $\Delta S^{\rm \, hier.}_k  (w_t) = S_{2^{\lfloor \log_2 k \rfloor}} (w_t) - S_{2^{\lfloor \log_2 k \rfloor-1}} (w_t)= - \ln p(w_t|w_{t-\lceil k/2 \rceil},\dots, w_{t-k})$ in the hierarchical access case, isolating the contribution of depth $\lfloor \log_2 k \rfloor$ in a hierarchical temporal model. This change of surprisal with context length gives us a re-factorised next-token likelihood which isolates contributions of past context to prediction refinement under a fictive autoregressive (sequential) or hierarchical temporal model. 

For each context access scheme, we construct a global log-transition probability matrix ($M\in R^{W \times L}$) with varying context length $1< L\leq 250$, constructed as $M_{t,k}=\Delta S_k (w_t)$. Note that, as the number of tokens is several orders of magnitude larger than the maximal context length ($W >\!\!> L$), the rank of the global transition matrix is constrained by the rank of its row space, giving a linear upper bound ($L$) on rank in the sequential access case and a logarithmic upper bound ($\lfloor \log_2L\rfloor$) in the hierarchical access case. 

\subsection*{Results}

Fig~\ref{fig:results}B demonstrates that both sequential and hierarchical context access follow a power-law decay in token-level surprisal: the amplitude of surprisal contribution decreases exponentially with context length, showing that tokens further in the past contribute significantly less to new predictions. However, sequential access has a smooth, monotonic decline in contribution, whereas hierarchical access produces a step-wise “staircase” profile.
This difference is reflected in the rank of the global transition matrix (Fig~\ref{fig:results}C). 
Interestingly, while the rank of the hierarchical context access follows its logarithmic upper bound, the rank in the sequential context access case is also logarithmic and thus increasingly deviates from its theoretical linear upper bound as context increases. This highlights that sequential context access is inefficient for iteratively refining predictions, e.g., using only $70/250 = 28\%$ of the maximal rank at context length $L=250$. Fig~\ref{fig:results}D shows that efficiency -- defined as the ratio of the amplitude of the surprisal contribution to the rank -- is consistently lower in hierarchical access than in sequential access.

\subsection*{General discussion}
\cite{parr2025beyond} frame transformers as autoregressive solutions to non-Markovian sequence modelling, contrasting them with deep temporal models that compress memory through hierarchical and temporal depth. Building on this, we show that structuring context access in iterative inference with transformer language models can mimic the surprisal factorisation of deep temporal models. Most notably, the hierarchical access to previous context, which naturally arises from inductive biases in deep temporal models, efficiently computes the contributions of previous context to iteratively refine predictions.  
This dissociates the model architecture, which underlies learning and representation, from the computations deployed for constructing predictions at inference time. This aligns with recent cognitive science studies~\citep{duncan2014associative, lake2017building,kwon2025coordinated,gershman2025key}, which suggest that inference processes in the brain can depart from their underlying model architectures.

\paragraph{Acknowledgements}
This work was funded by a gift from the Chan Zuckerberg Initiative Foundation to establish the Kempner Institute for the Study of Natural and Artificial Intelligence at Harvard University, and by the K. Lisa Yang Integrative Computational Neuroscience (ICoN) Center of the Yang-Tan Collective at MIT. 

\newpage
\small

\bibliography{ref.bib}
\bibliographystyle{abbrvnat}

\end{document}